\documentclass[iop]{emulateapj}

\usepackage {epsfig,graphicx,color}
\usepackage {txfonts   }
\usepackage {natbib    }
\usepackage {float     }
\usepackage {enumerate }
\usepackage {tabularx  }
\usepackage {url       }
\usepackage {capt-of   }
\usepackage {color     }

\begin{document}

%

\title{Fingering convection and cloudless models for cool brown dwarf atmospheres}

\author{ 
P. Tremblin\altaffilmark{1,2} and
D. S. Amundsen\altaffilmark{1}   and 
P. Mourier\altaffilmark{1,3}    and
I. Baraffe\altaffilmark{1,4}  and
G. Chabrier\altaffilmark{1,4} and
B. Drummond\altaffilmark{1}   and
D. Homeier\altaffilmark{4}    and
O. Venot\altaffilmark{5}
       }

\altaffiltext{1}{
  Astrophysics Group, University of Exeter, EX4 4QL Exeter, UK}
  
\altaffiltext{2}{
  Maison de la Simulation, CEA-CNRS-INRIA-UPS-UVSQ, USR 3441, Centre
  d'\'etude de Saclay, 91191 Gif-Sur-Yvette, France}

\altaffiltext{3}{
Master ICFP, D\'epartement de
Physique, Ecole Normale Sup\'erieure, 24 Rue Lhomond, 75005 Paris, France
}

\altaffiltext{4}{
  Ecole Normale Sup\'erieure de Lyon, CRAL, UMR CNRS 5574, 69364 Lyon
  Cedex 07, France}

\altaffiltext{5}{
  Instituut voor Sterrenkunde, Katholieke Universiteit Leuven, Celestijnenlaan 200D, 3001 Leuven, Belgium}

\email{tremblin@astro.ex.ac.uk or pascal.tremblin@cea.fr}

\begin{abstract}
This work aims to improve the current understanding of the
  atmospheres of brown dwarfs, especially cold ones with spectral type T
  and Y, whose modeling is a current challenge. Silicate and
  iron clouds are believed to disappear at 
  the photosphere at the L/T transition, but cloudless models fail to
  reproduce correctly the spectra of T dwarfs, advocating for the
  addition of more physics, e.g. other types of clouds or internal
  energy transport mechanisms.

We use a one-dimensional (1D) radiative/convective equilibrium code
  \texttt{ATMO} to investigate this issue. This code includes both
  equilibrium and out-of-equilibrium chemistry and solves consistently
  the $PT$ structure. Included opacity sources are H$_2$-H$_2$,
  H$_2$-He, H$_2$O, CO, CO$_2$,
  CH$_4$, NH$_3$, K, Na, and TiO, VO if they are present in the atmosphere.

We show that the spectra of Y dwarfs can be accurately reproduced
  with a cloudless model if vertical mixing and NH$_3$ quenching are
  taken into account. T dwarf spectra still have some
  reddening in e.g. $J-H$ compared to cloudless models. This reddening
  can be reproduced by slightly reducing the temperature gradient in the
  atmosphere. We propose that this reduction of the {\it stabilizing} temperature gradient in these layers, leading to cooler structures,
 is due to the onset of fingering convection,
triggered by the {\it destabilizing} impact of condensation of very thin dust.
\end{abstract}

\keywords{Methods: observational --- Methods: numerical --- brown dwarfs}

\maketitle

%
%

\section{Introduction}

Brown dwarfs with effective temperatures $T_\mathrm{eff}$ below $\sim$1000~K
(T and Y spectral types)
are of great interest to understand the physics of cool atmospheres
and pave the way for future studies of cool exoplanets with the James
Space Webb Telescope (JWST) and the European Extremely Large Telescope
(E-ELT). Detailed spectra can now be obtained in the near-infrared
(NIR) for these objects using e.g. the Hubble Space Telescope (HST),
or the Gemini Near-Infrared Spectrograph. 

The reddening of M/L dwarfs ($T_\mathrm{eff}$ $\approx$ 2000~K) is thought to
be the result of the condensation of silicate dusts and the subsequent
cloud formation \citep{Tsuji:1996vu,Chabrier:2000hq,Allard:2001fh}. At the L/T
transition ($T_\mathrm{eff}$ 
$\approx$ 1300~K), this reddening disappears with a sharp transition
as a function of $T_\mathrm{eff}$. Given the high amplitude
  variability observed at the L/T transition
  \citep[e.g.][]{Radigan:2014fg}, this effect is interpreted as the
clouds breaking up and forming holes at the photosphere
\citep[e.g.][]{Marley:2010kx}. Yet, the spectral modeling of T and Y dwarfs
remains a challenge because the IR colors predicted by the cloudless
models below  $\sim$1000~K 
appear too ``blue'' compared to observations. This additional
reddening has recently been interpreted by
\citet{Morley:2012io,Morley:2014gs}
as the emergence of sulfide and chloride clouds (MnS, Na$_2$S, and
KCl). However, even with this cloudy approach, it appears difficult to
obtain a good model of Y-dwarf spectra \citep[see for example WISEPC
  J1217b in][]{Leggett:2015dn}.

In the present paper, we show under which conditions the modeling
of T/Y-dwarf spectra is possible with a cloudless model. In
Sect.~\ref{sect:code}, we describe our 1D numerical code
\texttt{ATMO} and we apply it to T and Y dwarf atmospheres in
Sect.~\ref{sect:ty}. The spectral modeling of Y dwarfs is
in good agreement with observations when vertical mixing is taken into
account. For T dwarfs, good agreement is obtained when the temperature gradient
in the atmosphere is reduced. We discuss these results in
Sect.~\ref{sect:disc} and we 
suggest that the enhanced cooling in the deep layers of T dwarfs arises from the onset of fingering convection, resulting from the condensation
of some species in very thin dust. 

%
%

\section{Method and code description}\label{sect:code}

We have developed a 1D radiative/convective code \texttt{ATMO} solving
for the
pressure/temperature ($PT$) structure of an atmosphere, assuming a vertical
energy balance between the internal heat flux and the radiative and
convective fluxes, and vertical hydrostatic equilibrium. The
convection is taken into account using mixing length theory 
(MLT) as decribed in \citet{Henyey:1965hb} and \citet{Gustafsson:2008df} with a mixing length of
1.5 times the local pressure scale height.

The $PT$ structure is solved on a cartesian plane-parallel grid for a given effective
temperature $T_\mathrm{eff}$ and surface gravity $\log\,g$. The chemistry takes into
account $\sim$150 species at equilibrium with the
condensation of $\sim$30 liquids/solids assuming elemental abundances from
\citet{Caffau:2011ik}, and is solved by minimization of Gibbs free
energy. The references of the thermodynamic data for the gas-phase species can be found
in \citet{Venot:2012fr} except for H$_2$ whose thermodynamic data were
fitted from the Saumon-Chabrier equation of state \citep[see][]{Saumon:1995bu}.
The data for the condensed species can be found in
\citet{Robie:1968ud} and \citet{ChaseJr:1998ui}. In the present models, we neglect
rain-out processes and clouds. This approximation is critical
  to assess the abundances of some refractory elements such as Na, K, and
S because of the different condensation paths that can happen depending on the
thermal history of the atmosphere (e.g. NaAlSi$_3$O$_8$/KAlSi$_3$O$_8$
will not form if Al and Si are removed at higher temperatures and Na
and K will condense in Na$_2$S and KCl). However in the spectral
window of interest, the main absorbers such as
H$_2$O, NH$_3$, CH$_4$, CO are not affected by this hypothesis and
their abundances can even be computed analytically \citep[see][]{Burrows:1999gd}. The code is
 coupled to the kinetic chemical network of
\citet{Venot:2012fr}, which includes $\sim$1000 reversible
reactions associated with $\sim$100 species based on C, H, N, O
elements. Vertical mixing is taken into account with a free
parameter $K_{zz}$ representing the vertical eddy diffusion
coefficient \citep[see][]{Ackerman:2001gka}. We perform a full chemical
network run on the atmosphere by integrating the chemistry in time
(usually up to
10$^{10}$ seconds) and we reconverged the PT structure of the
atmosphere on the fly until both the chemistry and the PT structure
are in a steady state.
The
$PT$ structure can be solved consistently with the out-of-equilibrium
chemistry and can have a significant difference if a main absorber
is quenched below the photosphere by the vertical mixing. We
  tested the kinetic chemical network by reproducing the results of
  \citet{Venot:2012fr} for HD209458b and HD189733b. Our results are
  also in good agreement with \citet{Hubeny:2007hm}, especially
  considering the differences in elemental abundances and opacity
  line lists. For our case at 
  $T_\mathrm{eff}$=1000~K, log~g=4.5 and log~$K_\mathrm{zz}$=8.0, we predict a quenched
  abundance of CO at $\sim$2$\times 10^{-4}$ and of NH$_3$ at
  $\sim$7$\times 10^{-6}$, and the model with $T_\mathrm{eff}$=900~K,
  log~g=4.5 and log~$K_\mathrm{zz}$=8.0 in \citet{Hubeny:2007hm} indicates a CO
  abundance of $\sim 1.5\times 10^{-4}$ \citep[fast2 chemical
    timescale][]{Yung:1988bn} and 
  a NH$_3$ abundance of $\sim 5\times 10^{-6}$. 

The line-by-line radiative transfer has already been described and
tested in
\citet{Amundsen:2014df}. The differences in radiative flux and
  heating rate between our line by
  line code and the radiative scheme used in the global circulation
  model of the UK Met-Office, the Unifed Model \citep{Amundsen:2014df}
  are generally of the
  order of a few percent. We use the same opacities for H$_2$-H$_2$, H$_2$-He,
NH$_3$, H$_2$O, CO, TiO and VO \citep[see Table 1
  in][]{Amundsen:2014df}. Methane (CH$_4$) 
has been updated with the new line list from the Exomol project
\citep{Yurchenko:2014eb} and we included the CO$_2$ opacity using the line
list from \citet{Tashkun:2011es} with the line broadening available in
the literature (broadening coefficients from \citet{Thibault:1992jj,Thibault:2000hs,Sharp:2007kv,Padmanabhan:2014ho}).  We have also calculated opacities due to
the extremely pressure-broadened sodium and
potassium doublets using the line profiles of
\citet{Allard:2007cc}. We have used the correlated-$k$ method 
described in \citet{Amundsen:2014df} to
speed up the convergence and combined the $k$-coefficients by using
the random
overlap method for the mixture \citep{Lacis:1991uo}. Then we
  compute the surface spectrum from the converged PT structure using
  the line
  by line radiative transfer. The radiative
transfer module also includes Rayleigh scattering by H$_2$ and He,
using the accelerated $\Lambda$-iteration technique 
described in \citet{Bendicho:1995up}.

%
%


\begin{figure*}[t]
\centering
\includegraphics[width=0.49\linewidth]{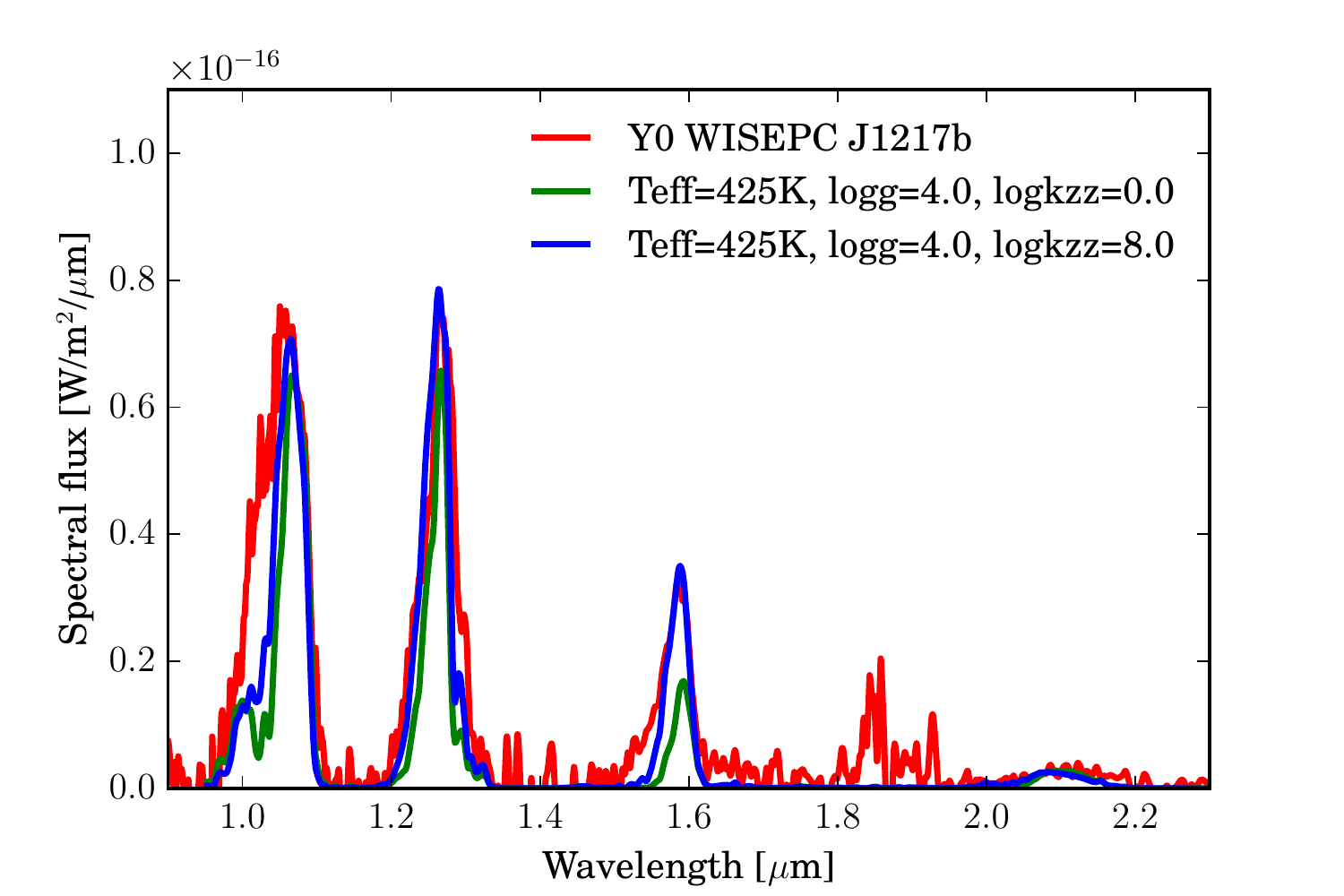}
\includegraphics[width=0.49\linewidth]{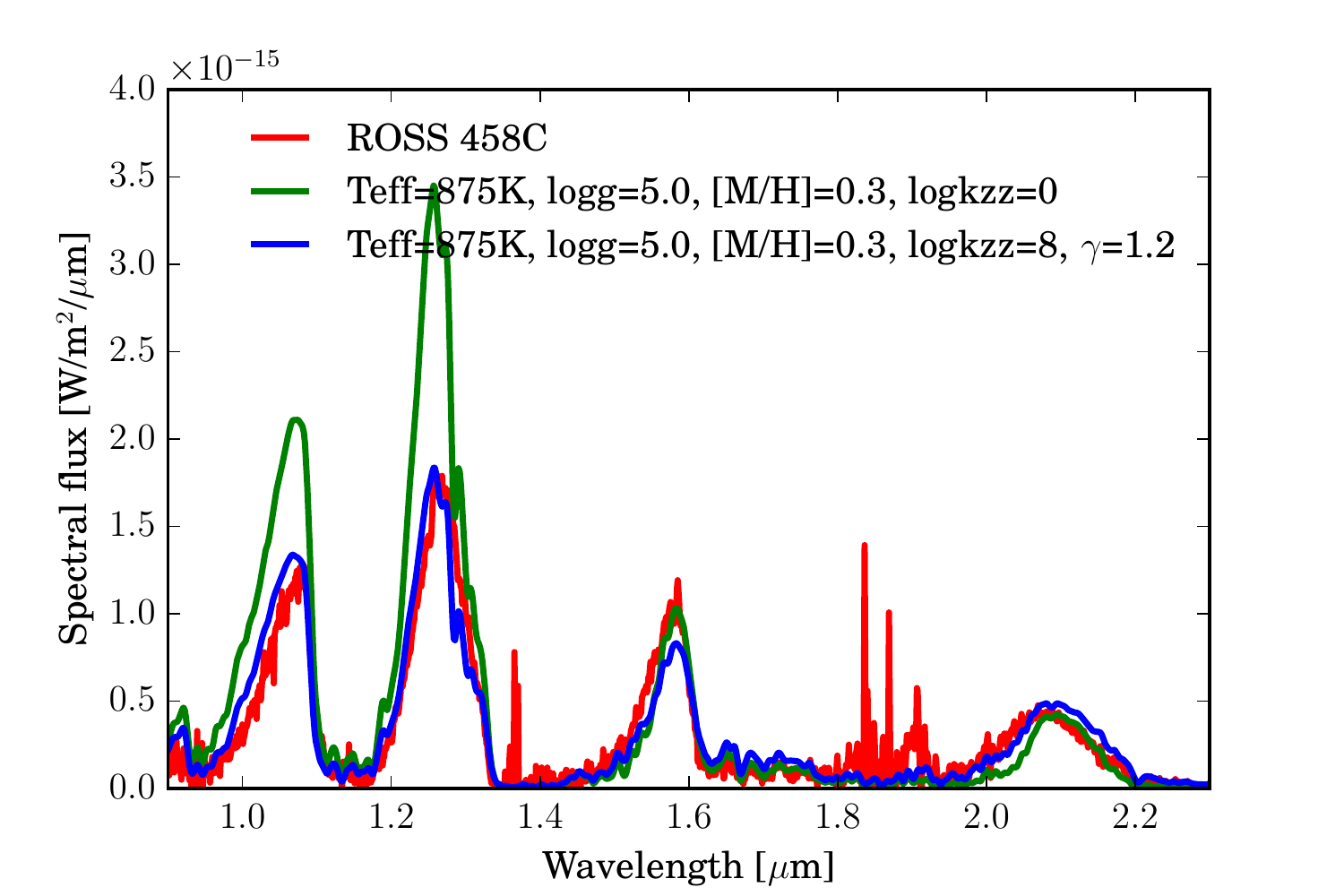}
\includegraphics[width=0.49\linewidth]{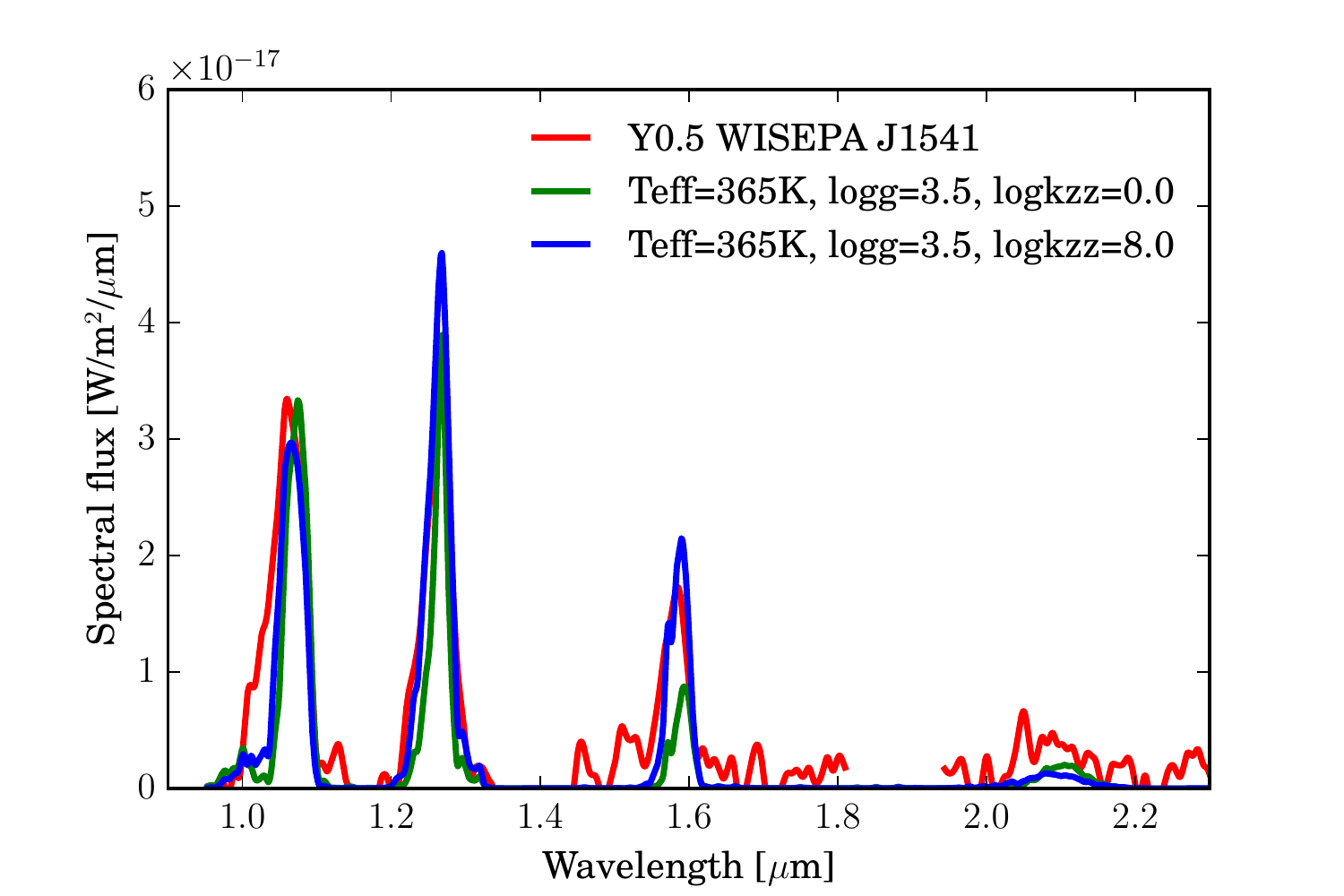}
\includegraphics[width=0.49\linewidth]{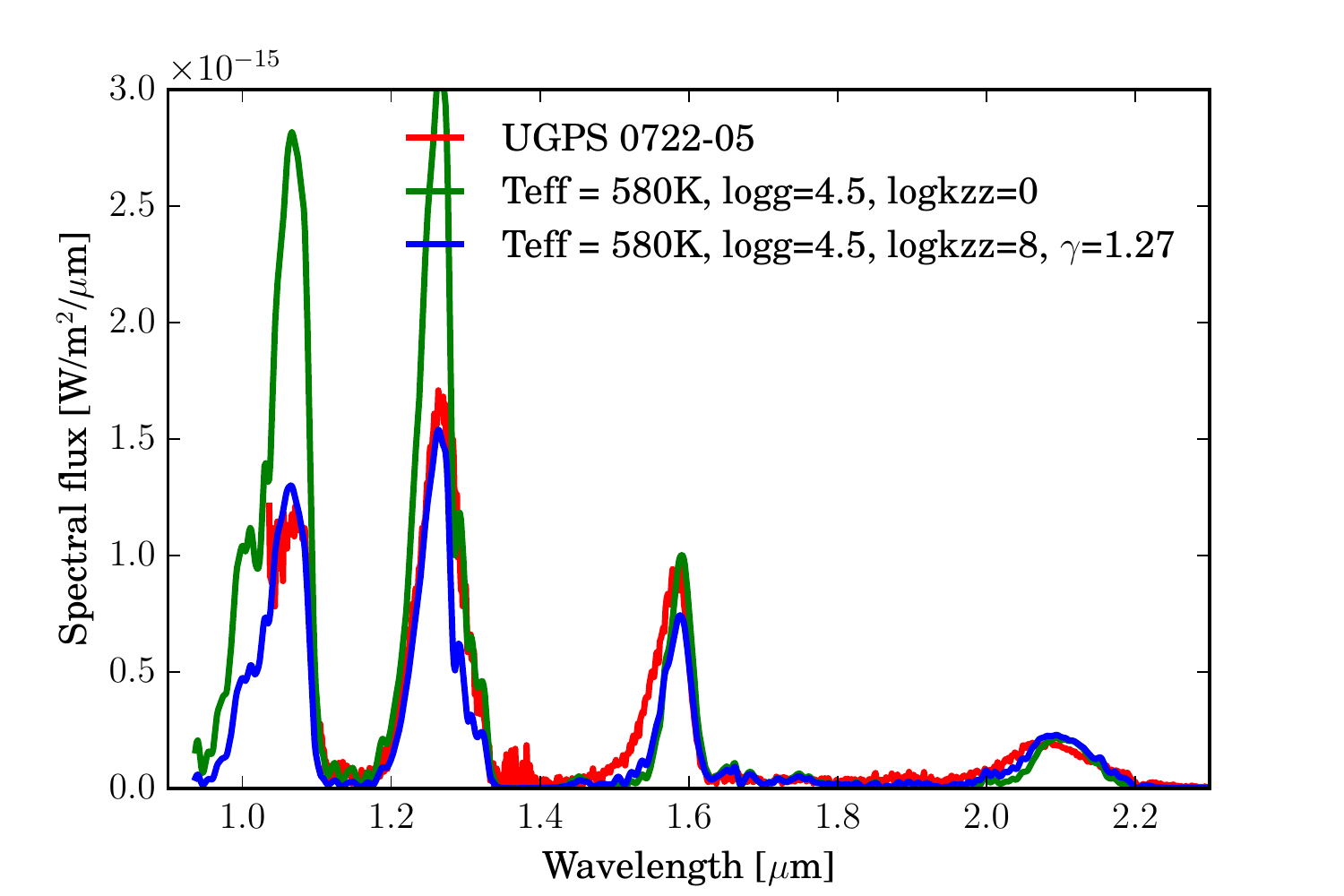}
\caption{\label{fig:spec} Left panel, top: spectral flux of WISEPC
  J121756.91+162640.2b (Y0) in red compared to a cloudless model with no vertical
  mixing (green)
  and with vertical mixing ($K_{zz}$ = 10$^8$ cm$^2$s$^{-1}$,
  blue). Left panel, bottom: similar
  models for the spectral flux of WISEPA J154151.66−225025.2 (Y0.5). The distances are taken from
parallax measurements: 10.1$^{+1.9}_{-1.4}$~pc \citep{Dupuy:2013ks}
and 5.7$^{+0.15}_{-0.14}$~pc \citep{Tinney:2014bl}, respectively.
Right panel, top: spectral flux of ROSS 458C in red
  compared to a cloudless model (green)
  and with a reduced temperature gradient (blue). Right panel, bottom: similar
  models for the spectral flux of UGPS J072227.51-054031.2. The
  distances are taken from 
parallax measurements: 11.7$^{+0.2}_{-0.2}$~pc
\citep{vanLeeuwen:2007dc} and 4.1$^{+0.6}_{-0.5}$~pc
\citep{Lucas:2010iq}, respectively. All the models
assume a radius of 0.1~R$_\odot$, no
adjustment of the total flux is used.}
\end{figure*}

\begin{figure}[t]
\centering
\includegraphics[width=\linewidth]{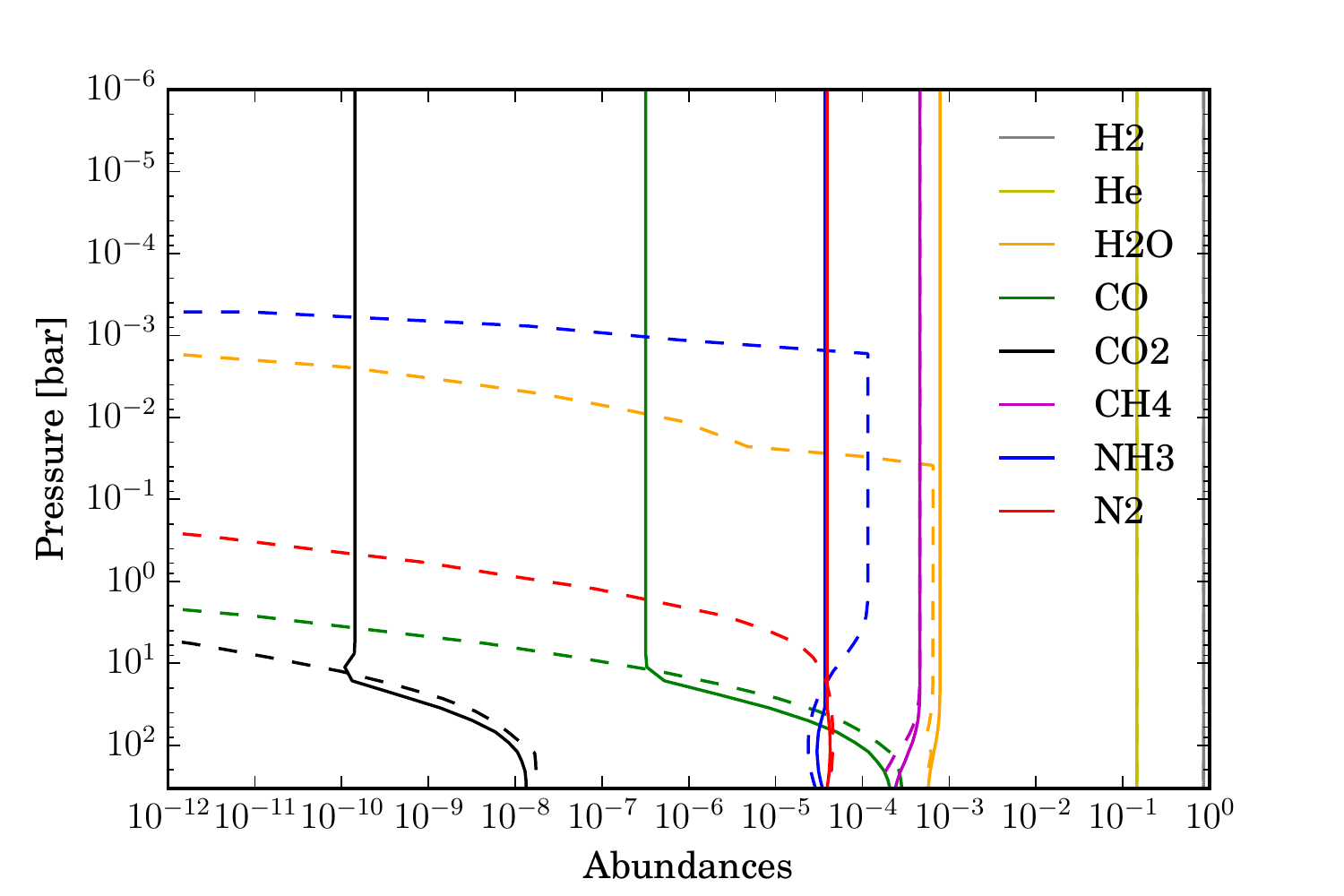}
\caption{\label{fig:NH3_quenching} Abundance profiles of the main
  species in the modeling of WISEPA J1541. The dashed profiles
  correspond to the model at chemical equilibrium and the solid
  ones to the model with vertical mixing with $K_{zz}$=10$^8$ cm$^2$s$^{-1}$.} 
\end{figure}

\section{Application to Y and T dwarfs}\label{sect:ty}

\subsection{Y dwarfs and NH3 quenching}\label{sect:yd}

\citet{Saumon:2006eo} have already demonstrated that ammonia is well
below its chemical equilibrium abundance in T dwarfs based on the
observed strength of th 9-11 $\mu$m absorption feature. It has also
later been confirmed in the NIR lines of NH$_3$
\citep[e.g.][]{Saumon:2012bl}, and \citet{Cushing:2011dk} confirmed
the signature of vertical mixing in all but one of their Y dwarfs by
the very absence of more detectable NH$_3$ absorption.
It has been expected that
absorption by ammonia in the $Y$ band should be present, but so far
the data have not identified such a signature 
\citep[e.g. Fig. 9 in][]{Cushing:2011dk}. It has been also
recently proposed by \citet{Leggett:2015dn} that ammonia is depleted
 in the atmospheres of Y dwarfs. Using the Y dwarf WISEPC
J1217b \citep[Y0,][]{Leggett:2014kz} as a template, we show in the
left panels of Fig.~\ref{fig:spec} that indeed a low
abundance of ammonia is required to reproduce the shape of the $Y$ band
flux around 1~$\mu$m and the correct flux in the $H$ band. 
Indeed, as shown in Fig.~\ref{fig:NH3_quenching}, when vertical mixing
is taken into account 
with a diffusion coefficient, the abundance profile of
ammonia is quenched at a relatively deep level (50-100 bars), resulting in a depletion of
ammonia by approximatively a factor 3, significantly reducing the
absorption in the peaks in the $Y$ band. We took a value of
$K_{zz}=10^8$~cm$^2$s$^{-1}$ for the eddy diffusion coefficient, which
is within the high values explored by \citet{Hubeny:2007hm}; however
the quenching of the NH$_3$/N$_2$ chemistry is relatively insensitive
to the choice of $K_{zz}$ \citep[see][]{Cushing:2011dk}. When we
include the vertical mixing, the condensation of H$_2$O and NH$_3$ is
ignored for simplicity, this approximation is not an issue because the
difference with and without condensation at equilibrium is small at
these effective temperatures ($T_\mathrm{eff}$ is shifted by $\sim$10~K). 
Importantly for the color-magnitude diagrams in the IR, 
the flux in the $H$ band at 1.6~$\mu$m is increased
and matches well the observed spectrum. As shown in
Fig.~\ref{fig:spec}, this increase in the flux is 
sufficient to produce the reddening required in $J-H$ to match the observations.
Applying the same physical treatment
to the cooler brown dwarf WISEPA J1541 \citep[Y0.5,][]{Cushing:2011dk}, we obtain the same significant improvement between observed and synthetic spectra,
as shown in the bottom-left panel of Fig.~\ref{fig:spec}. Since we have included the complete chemical network of
\citet{Venot:2012fr} in \texttt{ATMO}, we also predict the quenching of CO and
CO$_2$ (see Fig.~\ref{fig:NH3_quenching}), which will impact the
flux at 4.5 $\mu$m.

\subsection{T dwarfs and reduced temperature gradient}\label{sect:td}



\begin{figure}[t]
\centering
\includegraphics[width=\linewidth]{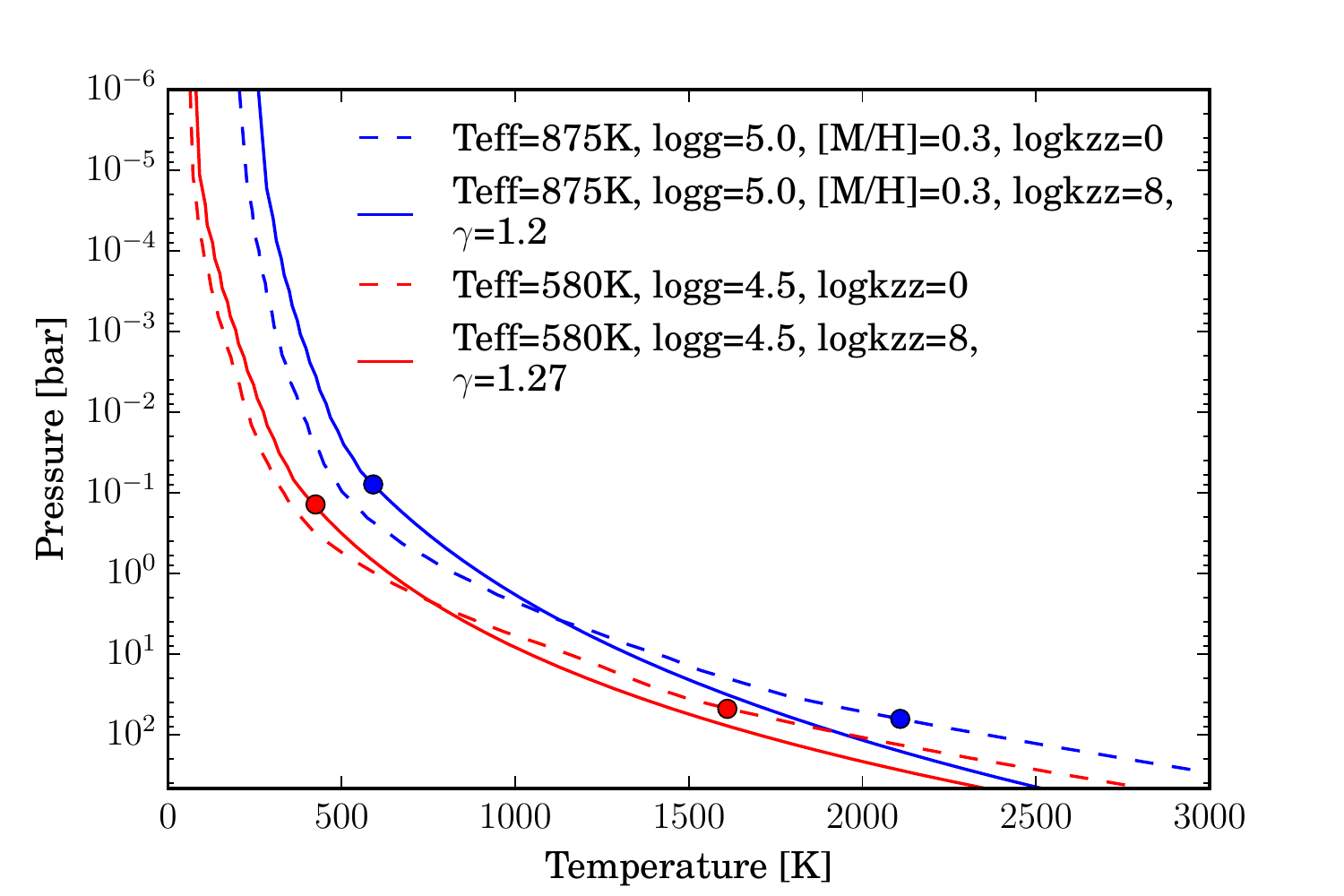}
\caption{\label{fig:pt_tdwarf} Pressure/temperature profiles of the
  models of ROSS 458C and UGPS 0722-05. The solid lines correspond to models
  with modified adiabatic index $\gamma$. The solid circles indicate the
  minimum pressures for which convection is present. Without a
  modified adiabatic index, the convection is overturning, while the
  modified index, illustrates heat transport by the fingering
  convection. The change in $K_\mathrm{zz}$ has little impact on the PT profile.}
\end{figure}

T dwarfs have bluer colors in $J-H$ and $J-K$ than L dwarfs which indicates that
the silicate and iron clouds should fall below the photosphere at the L/T
transition. However, their spectra still present a reddening compared
to cloudless models and \citet{Morley:2012io} proposed that another
type of clouds such as sulfide clouds might be responsible for this
residual reddening. We have calculated spectra for the T dwarfs ROSS 458C
\citep{Burgasser:2010bu} and UGPS 0722-05 \citep{Lucas:2010iq}
whose spectra are shown in right panels of Fig.~\ref{fig:spec}. As
suggested by \citet{Burgasser:2010bu}, we used a supersolar
metallicity ([M/H]=0.3) for ROSS 458C based on the measured
metallicity of its two companions ROSS 458 A and B.
Our cloudless
models overestimate the flux in the $Y$ and $J$ bands, yielding significantly
bluer colors in $J-H$/$J-K$ compared to observations. Since the
photosphere in the $Y$ and $J$ bands is deeper than the the photosphere in
the $H$ and $K$ bands, reducing the temperature gradient in the atmosphere
should decrease the $Y/J$ flux compared to the $H/K$ flux. We have tested this
idea by artificially modifying the adiabatic index $\gamma=C_P/C_V$ of the
atmosphere. The 
pressure/temperature profiles obtained with $\gamma$=1.2 for ROSS 458C and
$\gamma$=1.27 for UGPS 0722-05, respectively, are displayed in Fig.~\ref{fig:pt_tdwarf},
and the corresponding spectra in Fig.~\ref{fig:spec}. These models correctly reproduce
the spectra and the colors in $J-H$/$J-K$ without the need to invoke
additional cloud
opacity. As shown in Fig.~\ref{fig:pt_tdwarf}, the amplitude of the temperature reduction
required to match the observed spectra decreases with the effective temperature. Therefore, for Y dwarfs we found that there is no need to
adjust the adiabatic index with respect to its expected value ($\sim$~1.45). The relative importance of the effects of vertical mixing
and the reduction of the temperature gradient can be seen on the
color-magnitude diagrams in Fig.~\ref{fig:mag_col}.

%
%

\section{Discussion: Atmospheric fingering convection}\label{sect:disc}

\begin{figure}[t]
\centering
\includegraphics[width=\linewidth]{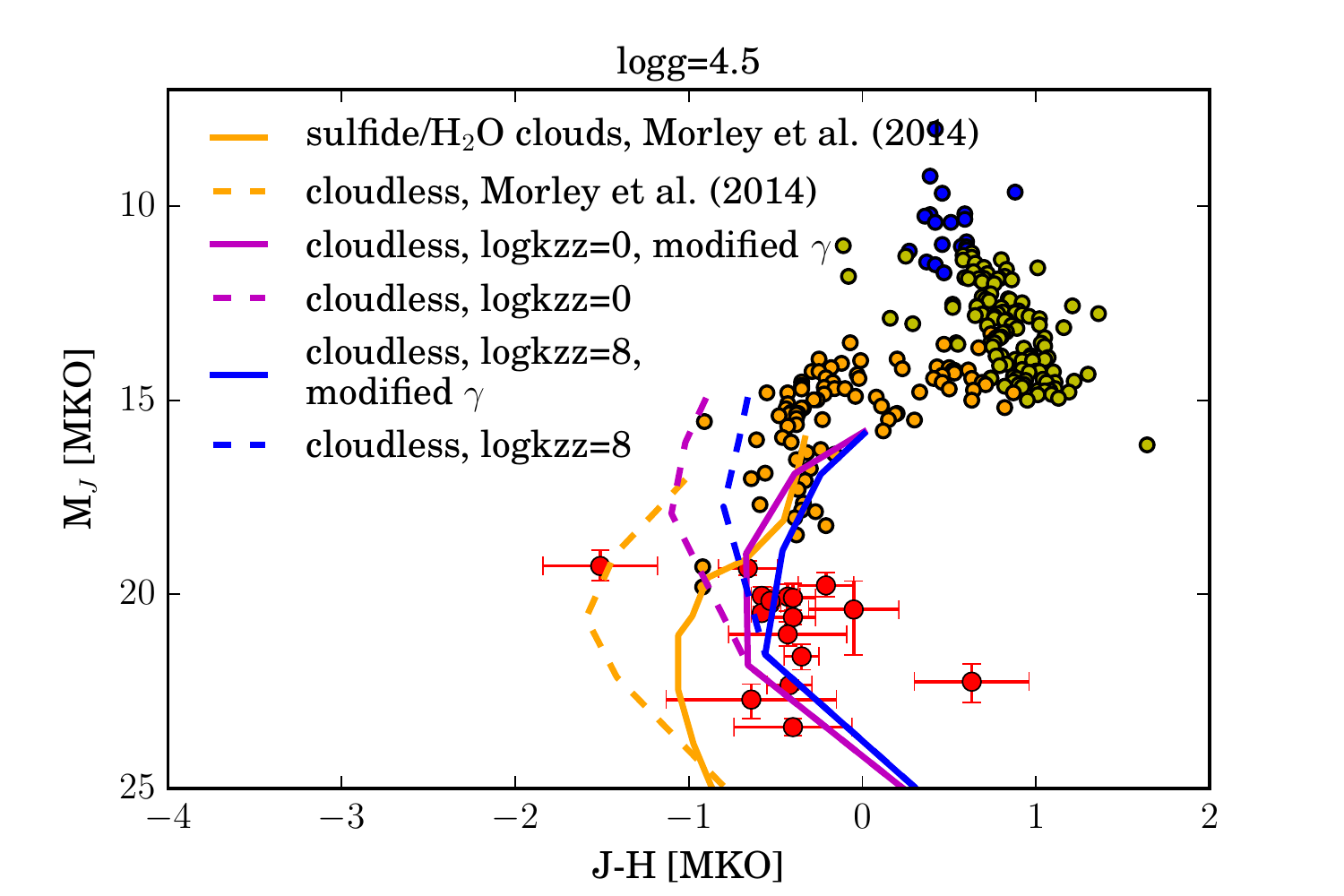}
\includegraphics[width=\linewidth]{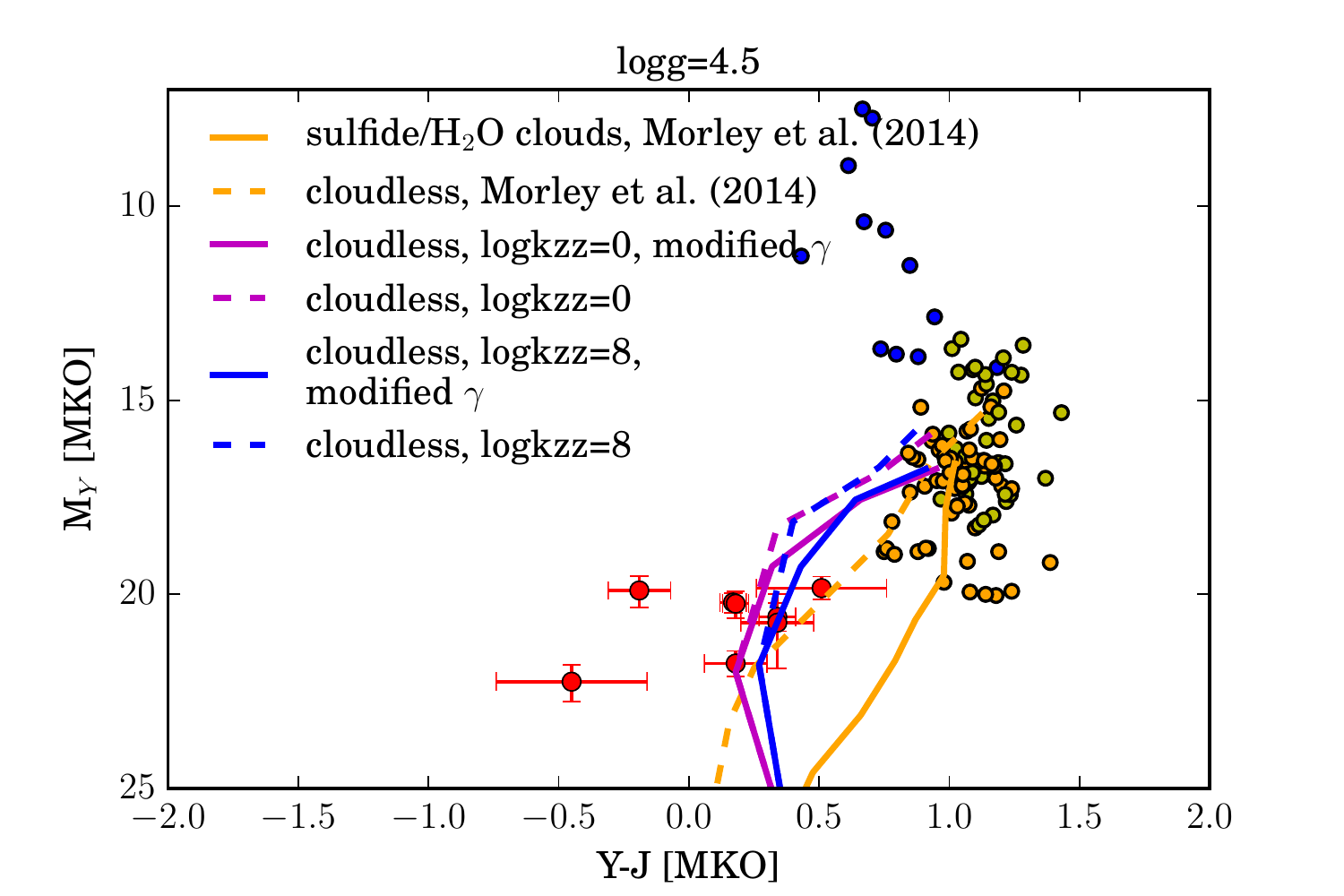}
\caption{\label{fig:mag_col} Color magnitude diagrams (top: J-mag vs
  $J-H$, bottom: Y-mag vs $Y-J$) for T/Y brown dwarfs. The Y-dwarf
  photometry is from \citet{Dupuy:2013ks,Beichman:2014jr} and the
  L/T/M from \citet{Dupuy:2012bp,Faherty:2012cy}. The modified
  adiabatic index is $\gamma$=(1.15,1.2,1.25) for
  $T_\mathrm{eff}$=(1000K,800K,600K), no modification for $T_\mathrm{eff}<$600K.} 
\end{figure}

As shown in Sect.~\ref{sect:yd}, a coherent cloudless model taking into account
vertical mixing in the atmosphere is able to reproduce the near-infrared spectrum of Y
dwarfs. Figure~\ref{fig:mag_col} shows the color-magnitude diagrams
for various cool T and Y brown dwarfs in $M_J$ vs $J-H$ and $M_Y$ vs $Y-J$. We have plotted
a series of different models at constant $\log\,g=4.5$ as well as the cloudless
and sulfide/H$_2$O cloud models of \citet{Morley:2014gs}. We display the
models with and without vertical mixing and with and
without modified adiabatic index to illustrate the relative impacts of
these processes. Although the cloudy models
do yield redder colors in $J-H$, providing a better match to the observed colors (see Fig.~16 in
\citet{Morley:2014gs}), they predict also a large
reddening in $Y-J$ which is incompatible with the observations. The
problem is even worse at larger gravity ($\log\,g=5$). In contrast, our cloudless models
reproduce the correct $J-H$ and $Y-J$ colors for Y dwarfs, without any
adjustable parameter. Furthermore, as shown in Sect.~\ref{sect:yd}, cloudless models consistently taking into account departure from chemical equilibrium due to vertical mixing 
in the atmosphere nicely reproduce the near-infrared spectrum of Y
dwarfs. Our cloudless models  are shifted in $J-H$ with
  respect to the ones of
  \citet{Morley:2014gs} by nearly one magnitude
  at log~g=4.5 and we would get similar offsets if we compare our models
  at log~g=4.5 to their models at log~g=5.0. Part of the shift (0.3 mag) can be
  explained by changing in our calculations, (i) the elemental abundances to
  \citet{Lodders:2003bf}, (ii) the line list for methane to the STDS line
  list \citep{Wenger:1998kk}, (ii) the thermodynamical data of
  H$_2$ to the one of the JANAF database as used in
  \citet{Morley:2014gs}. The origin of the rest of 
  the shift remains unknown but is probably due to differences in the
  opacities of the other molecules. It seems more
  likely, however, that such low-temperature  low-mass objects have gravities
  close to log~g=4.5 rather than log~g=5.0 \citep{Chabrier:2000ex}.

\begin{figure}[t]
\centering
\includegraphics[width=\linewidth]{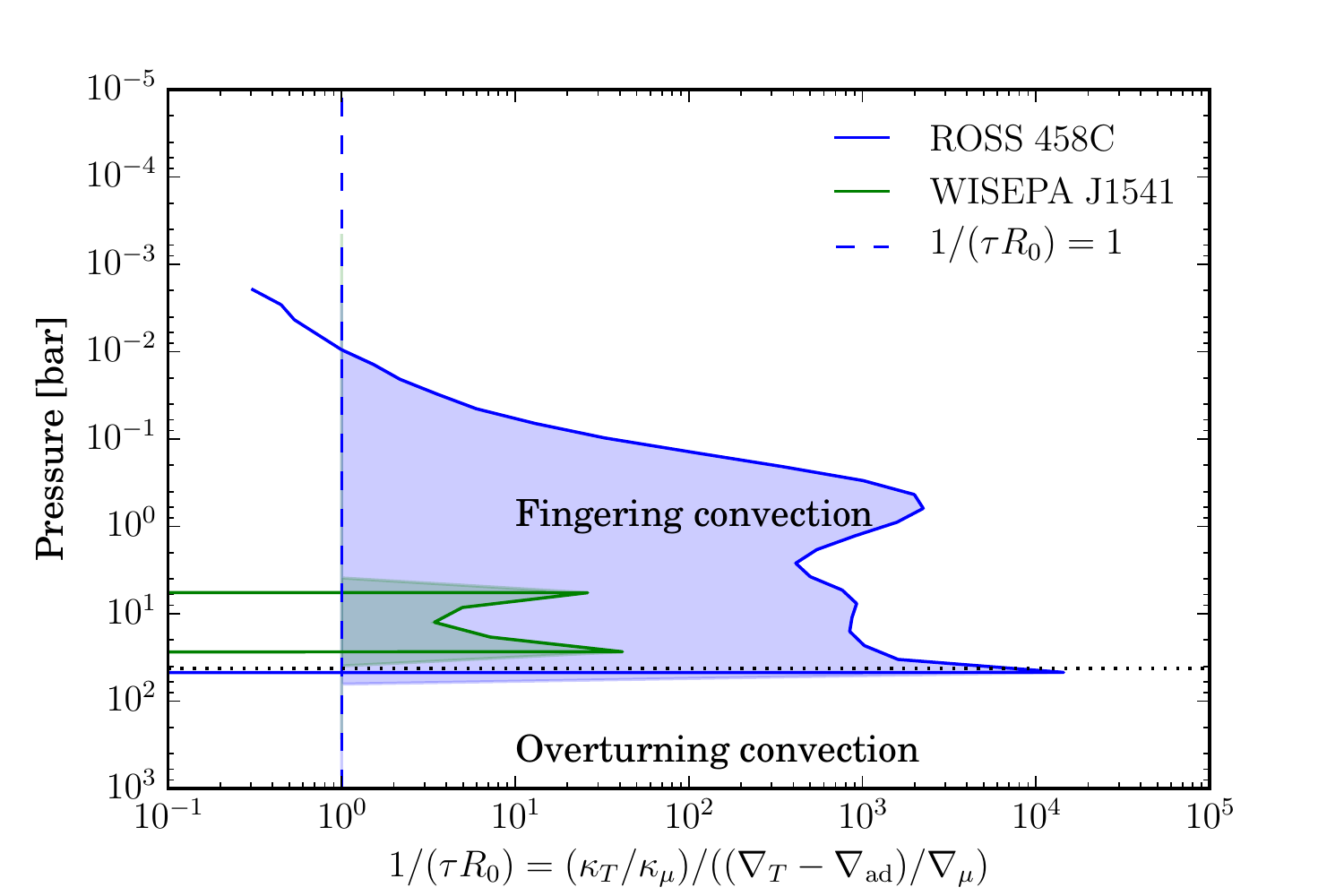}
\caption{\label{fig:finger_conv} Profile of $1/(\tau R_0)$ for ROSS 458C
  and WISEPA J1541. Fingering (thermohaline) convection is possible
  where $1/(\tau R_0)>1$. The dotted horizontal line shows the limit
  between the overturning and fingering convection zones for ROSS 458C.} 
\end{figure}

For T dwarfs, both our cloudless models and the cloudy models reproduce the
$J-H$ and $Y-J$ colors. In the former case, as shown in Sect.~\ref{sect:yd}, a good match of the spectrum requires a lower value of the atmospheric characteristic adiabatic
index (a good match is also obtained in $M_J$ versus $J-K$). This
adjustment can thus reflect either the effect 
of clouds or the signature of an energy transport mechanism that
reduces the temperature gradient in the atmosphere. Given the fact
that, for T dwarfs, iron and silicate clouds are expected to have settled below the phostosphere, as mentioned previously,
we favor the second suggestion. We suggest that the cooling is due to
the onset of fingering (thermohaline) convection in T dwarf deep atmospheric layers, similar to the one
 triggered by salt gradients in
Earth oceans \citep[e.g.][]{Rahmstorf:2003vg}. In the limit of ``infinitely-thin'' dust condensation, we have computed
the mean-molecular-weight gradient in the atmosphere,
$\nabla_\mu$\footnote{$\nabla_X = \mathrm{d}\ln(X)/\mathrm{d}\ln(P)$}, assuming
that all the condensates remain in the atmosphere in their
``mono-molecular'' form (e.g., assuming that equilibrium chemistry
predicts a given abundance of solid MgSiO$_3$, we compute the
mean molecular weight by keeping this abundance with the molecular
weight of MgSiO$_3$). This mean-molecular-weight gradient
should be fairly representative of $\nabla_\mu$ in the presence of small grains
if the grain formation process does not significantly change throughout the
atmosphere. As shown in \citet{Rosenblum:2011jb} \citep[see also][]{Leconte:2012gt}, the value of the dimensionless parameter $R_0=(\nabla_T-\nabla_\mathrm{ad})/\nabla_\mu$ 
determines the extent of the 
overturning convection zone 
($R_0<1$). In case the mean molecular weight increases with height
in a stable atmosphere ($\nabla_\mu<0$, and $(\nabla_T-\nabla_\mathrm{ad})<0$), as in the present context, 
the
possible existence of fingering convection is
determined by the relation $1<R_0<1/\tau$, where 
 $\tau=\kappa_\mu/\kappa_T$ defines the inverse Lewis number, i.e. the ratio of the molecular diffusivity
$\kappa_\mu$ to the radiative thermal
diffusivity $\kappa_T= 16\sigma T^3/(3\kappa \rho^2 C_p)$. 
This is the opposite situation to oscillatory double-diffusive convection, due to a {\it positive} molecular weight gradient and a destabilizing temperature gradient,
suggested to occur in some giant planet interiors \citep{Leconte:2012gt,Leconte:2013cx}.
Figure~\ref{fig:finger_conv} shows the atmospheric profiles of
$1/(\tau R_0)$ for ROSS 458C and WISEPA J1541, taking into account the
molecular diffusivity of MgSiO$_3$(s) and assuming 
that the thermal diffusivity is dominated by the radiative
processes ($\kappa_T$ varies from $\sim 10^{-1}$ cm$^2$s$^{-1}$ at 1
kbar to  $\sim 3\times 10^{5}$ cm$^2$s$^{-1}$ at 1
bar). As seen, fingering convection (i.e. $1/(\tau R_0)>1$)
is predicted to occur in some parts of these atmospheres. The
magnitude of the mean-molecular-weight gradient is modest:
$|\nabla_\mu|~\sim 10^{-4}-10^{-5}$ for ROSS 458C.
Interestingly, while the extension of the
fingering-convection zone is very large for ROSS 458C, it remains
relatively small
for the Y dwarf WISEPA J1541. This is fully consistent with the fact that we
need a lower adiabatic index for T dwarfs whereas this
modification is not necessary for Y dwarfs. 

The transport mechanism needed to lower the temperature
  gradient has to be efficient to transport energies at the level of
  the internal flux of the object. The possibility that fingering
  convection could carry out such energy fluxes remains to be
  demonstrated. Also the mixing induced by the convective fingers will reduce
  the destabilizing mean-molecular-weight gradient. A consistent
  theory of fingering convection in our model is needed in order to
  assess the possibility to have a self-sustained steady-state energy
  flux from the convection. Nevertheless our results strongly suggest
that fingering convection could occur in T dwarf atmospheres due to
the destabilizing effect of dust condensation in the stably stratified
part of the atmosphere,
potentially impacting the 
energy transport in these objects compared to both warmer and cooler objects and
leading to efficient cooling of their deep atmospheric layers. It can
be expected that the increase of the convective energy transport will
reduce the radiative flux in the corresponding region, hence reducing
the temperature gradient in the atmosphere, as needed for the
reddening in $J-H$.

\section{Conclusions}

In the present paper, we have shown that

\begin{itemize}
\item  Cloudless atmosphere models can reproduce the observed infrared Y
dwarf spectral energy distribution, provided vertical mixing
and out-of-equilibrium chemistry are properly taken into account to 
correctly predict the quenching of ammonia at deep levels, modifying its abundance profile.

\item Cloudless
models for T dwarfs with a reduced temperature
gradient in the atmosphere correctly reproduce observed fluxes and
colors. Such a reduced gradient can be obtained by
a modification of the adiabatic index in the atmosphere and could reflect
either the effect of clouds or of another type of energy transport in the
atmosphere. If clouds are effectively responsible for the reddening,
the modification of the adiabatic index is an easy way to mimic the
effect and could be used to better constrain cloud models that are
currently used. 

\item We suggest that fingering convection could be responsible for such a
reduced temperature gradient. We demonstrate that the condensation of
very thin dust under typical
T dwarf atmosphere conditions could trigger this mechanism and that the extent of this process
decreases with decreasing effective temperature, essentially vanishing for Y dwarf atmosphere conditions. 
\end{itemize}

Fundamental physical mechanisms such as atomic diffusion and
hydrodynamical instability might thus take place in cold brown dwarf atmospheres and play a major 
role in their spectral
evolution. Future observations of cool objects with SPHERE, GPI, and JWST
combined with 
  comparisons with the different models 
   should enable us to distinguish between these two effects: presence of clouds
  or reduced temperature gradient in T dwarf atmospheres. Our
  ability to constrain which physics is indeed present in cool brown
  dwarf atmospheres will bear important consequences for the future understanding of cool
  exoplanet spectra.

%
%

\begin{acknowledgements}
We thank Adam Burgasser, Sandy Leggett, Philip Lucas, Mike Cushing, and
Chas Beichman for providing their data. The 
calculations for this paper were performed on the DiRAC Complexity 
machine, jointly funded by STFC and the Large Facilities Capital Fund
of BIS, and the University of Exeter Supercomputer, a DiRAC Facility
jointly funded by STFC, the Large Facilities Capital Fund of BIS, and
the University of Exeter. This work is partly supported by
the European Research Council under the European Community's Seventh
Framework Programme (FP7/2007−2013 Grant Agreement No. 247060). 
Part of this work is supported by the Royal Society award WM090065
and the consolidated STFC grant ST/J001627/1. O.V. acknowledges
support from the KU Leuven IDO project IDO/10/2013 and from the FWO
Postdoctoral Fellowship programme. 
\end{acknowledgements}

%
%



\end {document}